

\magnification 1200
\overfullrule 0 pt






\def\CcC{{\hbox{\tenrm C\kern-.45em{\vrule height.65em width0.07em depth-.04em
\hskip.45em }}}}
\def\RrR{{\hbox{\tenrm I\kern-.17em{R}}}}
\def\HhH{{\hbox{\tenrm {I\kern-.18em{H}}\kern-.18em{I}}}}
\def\DdD{{\hbox{\tenrm {I\kern-.18em{D}}\kern-.36em {\vrule height.62em 
width0.08em depth-.04em\hskip.36em}}}}
\def\ZzZ{{\hbox{\tenrm Z\kern-.31em{Z}}}}
\def\IiI{{\hbox{\tenrm I\kern-.19em{I}}}}
\def\NnN{{\hbox{\tenrm {I\kern-.18em{N}}\kern-.18em{I}}}}
\def\QqQ{{\hbox{\tenrm {{Q\kern-.54em{\vrule height.61em width0.05em 
depth-.04em}\hskip.54em}\kern-.34em{\vrule height.59em width0.05em depth-.04em}}
\hskip.34em}}}
\def\OoO{{\hbox{\tenrm {{O\kern-.54em{\vrule height.61em width0.05em 
depth-.04em}\hskip.54em}\kern-.34em{\vrule height.59em width0.05em depth-.04em}}
\hskip.34em}}}

\def\uq2{U_q({\uit su}(2))}

\def\fraz#1#2{{\strut\displaystyle #1\over\displaystyle #2}}

\def\part#1{\fraz{\partial}{\partial#1}}

\def\ii#1{\item{$\phantom{1}#1~$}}

\def\su2q{SU(2)_q}
\def\h1q{H(1)_q}

\def\nu{N_{1}}

\hsize= 15 truecm
\vsize= 22 truecm
\hoffset= 0.5 truecm
\voffset= 0 truecm
 
\null\vskip1.5truecm
 
\baselineskip= 13.75 pt
\footline={\hss\tenrm\folio\hss} 
\centerline 
{\bf  QUANTUM GROUPS, COHERENT STATES, SQUEEZING}
\smallskip
\centerline
{\bf AND LATTICE QUANTUM MECHANICS}
\bigskip
\centerline{
{\it    E.Celeghini ${}^1$, S.De Martino ${}^2$, S.De Siena ${}^2$,
	     M.Rasetti ${}^3$ and G.Vitiello ${}^2$.}}
\bigskip
{\it ${}^1$Dipartimento di Fisica - Universit\`a di Firenze and
INFN--Firenze, I 50125 Firenze, Italy}
 
{\it ${}^2$Dipartimento di Fisica - Universit\`a di Salerno and INFN--Napoli,
I 84100 Salerno, Italy}
 
{\it ${}^3$Dipartimento di Fisica and Unit\`a INFM -- Politecnico di 
Torino, I 10129 Torino, Italy}
\bigskip
\bigskip
\bigskip
\bigskip
\bigskip
\bigskip
\bigskip
 
\centerline{\bf Abstract} 
\medskip 
\noindent{ By resorting to the Fock--Bargmann
representation, we incorporate the quantum Weyl--Heisenberg algebra, $q$-WH,
into the theory of entire analytic functions. The $q$--WH algebra operators are
realized in terms of finite difference operators in the $z$ plane. In order to
exhibit the relevance of our study, several applications to different cases of
physical interest are discussed: squeezed states and the relation between
coherent states and theta functions on one side, lattice quantum mechanics and
Bloch functions on the other, are shown to find a deeper mathematical
understanding in terms of $q$-WH. The r\^ole played by the finite difference
operators and the relevance of the lattice structure in the completeness of the 
coherent states system suggest that the quantization of the WH algebra is an
essential tool in the physics of discretized (periodic) systems.}


\vfill
\noindent Annals of Physics (N.Y.), in press.
\eject
 
\bigskip
\noindent {\bf 1. Introduction}
\bigskip 
 
A great deal of attention and efforts have been recently devoted in theoretical
physics to the mathematical structures referred to as
$q$-groups$^{[1],[2],[3]}$. 
 
The basic features of these structures appear now pretty well understood, and
promise to be very rich of physical meaning, even though some properties still
deserve more study to be fully under control. The interest in $q$-groups arose
almost simultaneously in statistical mechanics as well as in conformal
theories, in solid state physics as in the study of topologically non-trivial
solutions to nonlinear equations, so that as a matter of fact, the research in
$q$-groups grew along parallel lines from physical as well as from mathematical
problems. 
 
The dual structures to $q$-groups, are the so called $q$-algebras: actually by 
a now almost universally adopted convention, the name $q$-groups designates
$q$-algebras as well, which are instances of Hopf algebras$^{[4]}$.
 
$q$-algebras are deformations in the enveloping algebras of Lie algebras for
which, contrary to the latter, the relevant Hopf algebra features are highly
non-trivial, whose structure  appears to be an essential tool for the
description of composed systems. The general properties of $q$-algebras are
better known than those of $q$-groups, in particular for the specific
characteristics which relate them with concrete physical models, due to the
complexity arising from the $C^*$-algebra properties of $q$-groups. 
 
Among the $q$-algebras, the non-semisimple ones are less easy to handle, as in
fact it happens in general with non-semisimple structures. We shall here focus
our attention on the $q$-deformation of the Weyl-Heisenberg algebra. The WH
algebra, admits two inequivalent deformations: one which is properly a
$q$-algebra$^{[5]}$, the other (on which we shall focus here our attention,
denoting it as $q$-WH, often referred to in the mathematical literature as
$osp_q (2|1)$), originated by the seminal work of Biedenharn$^{[6]}$ and
MacFarlane$^{[7]}$, which is characterized by the property that the intrinsic
nature of superalgebra -- proper also to the WH algebra itself --  plays a
non-trivial role, in view of the form of the coproduct.  It can therefore be
referred to as a Hopf superalgebra$^{[8], [9]}$. 
 
The $q$-WH algebra has had -- in analogy with the non-deformed case -- 
several applications in physics, such as the $''$quantum$''$ harmonic 
oscillator, $q$-coherent and squeezed states, Jordan-Wigner realizations, 
and so on. 
 
In this paper we cast the study of the $q$-WH algebra in the frame of the
Fock-Bargmann representation (FBR) of Quantum Mechanics (QM). We show that the 
analytic structure of the Lie algebra is fully preserved, and we can therefore
operate in a scheme where analyticity is ensured. 
 
As a first step we present a realization of the $q$-WH algebra in terms of
finite difference operators. As a result of this we are led to recognize that
whenever a finite scale is involved in a self-contained physical theory, a
$q$-deformation of the algebra of dynamical observables occurs, with the
$q$-parameter related with the finite spacing, namely carrying the information
about discreteness. $q$-deformation is also expected in the presence of
periodic conditions, since periodicity is but a special form of invariance
under finite difference operators. 
 
The analytic properties of the FBR together with the von Neumann lattice
topological structure in the complex plane, make then the relation between the
$q$-WH algebra and the coherent states (CS) in QM manifest. Here, we obtain a
formal relation between the coherent state generator and the commutator of
$q$-WH creation and annihilation operators, which is thus recognized to be an
operator in the CS space. Moreover, theta functions, in terms of which CS are
expressed, also admit a representation in terms of $q$-deformed WH commutators
on the von Neumann lattice, allowing us to get further insight in the basic
unity of the various structures. 
 
One additional successive step in the understanding of the physical meaning of
$q$-deformations is then achieved by realizing that the commutator of $q$-WH
creation and annihilation operators acts, in the FBR, like the squeezing
generator for CS, a result which confirms a conjecture, previously$^{[10]}$
formulated, whereby $q$-groups are the natural candidates to study squeezed
coherent states. 
 
Establishing a relation between CS and the $q$-WH algebra is of course of great
interest in view of the numerous interesting physical applications of the CS
formalism, and it may also open rich perspectives in Quantum Field Theory where
the CS formalism is the key to study vacuum structure and boson condensation. 
 
The relevance of $q$-deformation to discretized system physics naturally leads
us to analyze the structure of Lattice Quantum Mechanics (LQM). We study it in
configuration space as well as in momentum space and show that LQM is
characterized in both cases by the algebra $E(2)$. We complete the analysis of
LQM by presenting the lattice CS, optimizing the lattice position-momentum
uncertainty relation. $q$-WH turns out to be the algebraic structure underlying
the physics of lattice quantum systems. We find that the commutator between
$q$-WH creation and annihilation operators acts as generator of the $U(1)$
subgroup of $E(2)$, giving rise to phase variations in the complex plane. In
this context, in the presence of a periodic potential on the lattice, there
emerges naturally a relation between the $q$-WH algebra and the Bloch
functions, which further confirms the conjectured deeply rooted presence of
$q$-deformation within the dynamical structure of periodic systems. 
  
As a general remark, we would like to stress that it is only by fully
exploiting the FBR that we succeed in incorporating $q$-deformation of the WH
algebra into the theory of (entire) analytical functions. Such result may
deserve by itself further attention: in this way, indeed, it appears possible
to elucidate the deep r\^ole of $q$-WH algebra in the physics of lattice
quantum systems, coherent states and squeezing. 
 
One result of our analysis which also is remarkable and which will be widely
used in the following sections is the one which shows that the action of linear
operators of $q$-WH algebra may be represented in the FBR as the action of
nonlinear operators realized in FBR, e.g. the CS displacement operator. 
 
In this paper we do not pursue to its full extent the study of the r\^ole
played by the complete Hopf (super-)algebra structure of $q$-WH, that we expect
to enter into play when the analysis described here is extended to more degrees
of freedom.  In particular, we believe the notion of co-product will provide
the way of encompassing in a unique structure the non-trivial topological
features proper to the extension to higher dimension of theta functions, and
hence to several degrees of freedom of coherent states. Such an analysis needs
further and deeper formal investigation, which goes beyond the task of present
paper (concerned mainly with the possibility of relating the $q$-deformation
parameter with some physical quantity) and which we plan for future work. 
  
Through this paper we shall use units such that all relevant physical
quantities are dimensionless.

\bigskip
\noindent {\bf 2. $q$-Weyl-Heisenberg algebra,  Fock--Bargmann
representation and finite difference operators}
\bigskip
 
The Weyl-Heisenberg algebra is generated by the set of 
operators $\{ a, a^{\dagger}, \IiI \}$ with commutation relations 
$$
[ a, a^\dagger ] = \IiI \quad ,\quad [ N, a ] = - a \quad ,\quad [ N, 
a^\dagger ] = a^\dagger \quad , \eqno{(2.1)} 
$$
where $N \equiv a^{\dagger} a$. The Fock space of states ${\cal K}$ is nothing
but the representation of (2.1) generated by the eigenkets of $N$ with integer
(positive and zero) eigenvalues. Any state vector $\displaystyle{|\psi >}$ in
${\cal K}$ is thus described by the set $\{c_n|c_n\in \CcC\}$ defined by
$\displaystyle{|\psi > \, = \sum_{n=0}^\infty c_n |n>} \,$, {\it i.e.} by its
expansion on the complete orthonormal set of eigenkets $\{ |n> \}$ of $N$. 
 
The intrinsic nature of superalgebra of such scheme is manifestly shown by
noticing that, upon defining $H \equiv N + {1\over 2}$, the three operators 
$\{ a, a^{\dagger}, H \}$ close, on ${\cal K}$, the relations 
$$ 
\{ a, a^\dagger \} = 2 H \quad ,\quad [ H, a ] = - a \quad ,\quad [ H, 
a^\dagger ] = a^\dagger \quad , \eqno{(2.2)} 
$$
that are, once more on ${\cal K}$, equivalent to (2.1). 
 
We shall show in the sequel that the physical interpretation of {\sl quantum}
algebraic structures in the frame of discrete systems stems out, quite
naturally, from the quantization of the latter form, which -- contrary to the
quantization of (2.1) of ref. [5], which preserves the nature of algebra --
preserves the nature of superalgebra. 
 
The quantum (deformed) version of (2.2), $q$-WH, is, in terms of the set of
operators $\{ a_q, {\bar a}_q, H ;  ~q \in \CcC \}$ (where $H$ is assumed to be
the same as in (2.2))$^{[8]}$$^{[9]}$: 
$$
\{ a_q , {\bar a}_q \} = [ 2H ]_{\sqrt{q}} \quad ,\quad [ H , {\bar a}_q ] =
{\bar a}_q \quad,\quad [ H , a_q ] = - a_q  
\quad , \eqno(2.3) 
$$ 
where we utilized the customary notation 
$$
[ x ]_q \equiv {{q^{{1\over 2}x}-q^{-{1\over 2}x}}\over{q^{1\over 2} - 
q^{-{1\over 2}}}} \; . 
$$ 
 
The $q$-WH structure defined by (2.3) together with the related coproduct
$$
\eqalign{ 
\Delta(H) = H \otimes \IiI + \IiI \otimes H  \quad
\Rightarrow  
\quad   
\Delta(N) = N \otimes \IiI + \IiI \otimes N + \textstyle{{1\over 2}} 
\IiI \otimes \IiI \quad &,\cr   
\Delta(a_q) = a_q \otimes q^{{1\over 4} H} + q^{-{1\over 4} H} 
\otimes a_q \; \quad , \; \quad
\Delta({\bar a}_q) = {\bar a}_q \otimes q^{{1\over 4} H} + q^{-{1\over 4} H} 
\otimes {\bar a}_q 
\quad &, \cr}  
\eqno (2.4)
$$
is a quantum superalgebra (graded Hopf algebra) and, consequently, all
relations (2.3) are preserved under the coproduct map. 
 
In the space ${\cal K}$ (${\it i.e.}$ in the space spanned by the vectors
$\{|n> |~ n\in \NnN \}$), eqs.(2.3) can be rewritten in a more familiar
equivalent form$^{[6],[7],[9]}$, which makes them more explicitly analogous to
the undeformed case: 
$$
a_q {\bar a}_q - q^{-{1\over 2}} {\bar a}_q a_q = q^{{1\over 2}N} ,\quad\quad
[ N , a_q ] = - a_q \quad ,\quad [ N , {\bar a}_q ] = {\bar a}_q ;
\quad \eqno(2.5) 
$$ 
or, by introducing ~${\hat a}_q \equiv {\bar a}_q q^{N/2}$~,
$$
[ a_q, {\hat a}_q ] \equiv a_q {\hat a}_q - {\hat a}_q a_q = q^N ,\quad\quad
[ N, a_q ] = - a_q \quad ,\quad [ N, {\hat a}_q ] = {\hat a}_q \quad . 
\eqno(2.6)
$$
 
It must be stressed that (2.1) is a true Hopf algebra, whereas (2.5) and (2.6)
are only deformations at the algebra level of (2.1). Thus (2.3-4) is the
relevant mathematical structure of our problem.  We prefer however to resort
henceforth  to (2.6) -- even though the whole discussion which will follow
could, in principle, be based on (2.3) --  since it is perfectly correct as far
as we remain in ${\cal K}$ (and in its representations) and it is the most
similar to the usual form (2.1) of the WH algebra. 
 
$q$ will be assumed to be any complex number, except in Sect. 5, where it will
be restricted to be unimodular. The notion of hermiticity for the generators of
$q$-WH associated with complex $q$ is non trivial and has been studied in ref.
[10] in connection with the discussion of the feature of squeezing of the
generalized coherent states $(GCS)_q$ over ${\cal K}\, ^{[10]}$. 
 
The basic point of this paper is the functional realization of eqs. (2.6) by
means of finite difference operators in the complex plane, in the Fock-Bargmann
representation (FBR) of Quantum Mechanics$^{\, [11],[12]}$. In the FBR state
vectors are described by entire analytic functions, contrary to the usual
coordinate or momentum representation, where no condition of analyticity is
imposed. 
 
The Fock-Bargmann representation of the commutation relations (2.1) is: 
$$
N \to z {d\over dz} \quad ,\quad a^\dagger \to z \quad ,\quad a \to 
{d\over dz} \quad .\eqno{(2.7)}
$$
The corresponding eigenkets of $N$ (orthonormal under the usual gaussian
measure $d\mu(z)=$ $\displaystyle{{1\over{\pi}} e^{- |z|^2} dz d{\bar z}}$) are
$$
u_n(z) = {z^n\over \sqrt{n!}} \quad ,\quad  u_0 (z) = 1 \quad\quad
\quad\quad (n\in \NnN _+) \; . 
\eqno(2.8)
$$
 
The FBR is the Hilbert space generated by the $u_n(z)$, {\it i.e.} the whole
space  ${\cal F}$ of entire analytic functions. To each state vector $|\psi >$
is associated, in a one-to-one way, a function $\psi (z) \in {\cal F}$ by: 
$$
|\psi >  = \sum_{n=0}^\infty c_n |n> \quad\quad \rightarrow \quad\quad
\psi (z) = \sum_{n=0}^\infty c_n u_n(z) \; .
\eqno{(2.9)}
$$
Note also that, as expected in view of the correspondence ~${\cal K} \to
{\cal F}$~ (induced by $~|n> \to u_n(z)$),
$$\eqalign{
a^\dagger ~u_n (z) ~=~ \sqrt{n + 1} ~u_{n+1} (z) \quad &, \quad
a ~u_n (z) ~=~ \sqrt{n} ~u_{n-1} (z) \quad , \cr
N ~u_n (z) ~=~ a^\dagger a ~u_n (z) ~&=~ z {d\over dz} ~u_n (z) ~=~
n~ u_n (z) \; .\cr}
\eqno{(2.10)}
$$
(2.10) establish the mutual conjugation of $a$ and $a^\dagger$ in 
the FBR, with respect to the measure $d\mu (z)$. 
 
Let us now consider the finite difference operator ${\cal D}_q$ defined by: 
$$
{\cal D}_q f(z) = {{f(q z) - f(z)}\over {(q-1) z}} 
\quad , \eqno{(2.12)}
$$
with ~$f(z) \in {\cal F}\; ,\; q = e^\zeta \; ,\; \zeta \in {\CcC}$ . ${\cal
D}_q$ is the so called $q$-derivative operator$^{\, [13]}$, which, for $q \to
1$ ($\zeta \to 0$), reduces to the standard derivative. By using (2.8) and
(2.10), it may be written on ${\cal F}$ as 
$$
{\cal D}_q = \bigl((q-1) z \bigr)^{-1} 
\Bigl( q^{z{d\over {dz}}}  - 1\Bigr) 
=  q^{{z\over 2}{d\over {dz}}} {1\over z}\, \left [ z {d\over {dz}} 
\right ]_q \quad . 
\eqno{(2.13)}
$$
 
Consistency between (2.12) and the latter form of ${\cal D}_q$ can be proven 
by first $''$normal ordering$''$ the operator ${\left ( z {d\over{dz}} \right
)^n}$ in the form: 
$$
\left ( z {d\over{dz}} \right )^n = \sum_{m=1}^n {\cal S}_n^{(m)} z^m 
{{d^m}\over{dz^m}} \quad , \eqno{(2.14)}
$$
where ${\cal S}_n^{(m)}$ denotes the Stirling numbers of the second 
kind, defined by the recursion relations$^{\, [14]}$   
$$
{\cal S}_{n+1}^{(m)} = m ~{\cal S}_n^{(m)} + {\cal S}_n^{(m-1)} \quad , 
\eqno{(2.15)} 
$$
and then expanding in formal power series the exponential $\displaystyle{ 
\left ( q^{z {d\over {dz}}} - 1 \right )}$, keeping in mind the identity:
$$ 
{1\over{m!}} \left ( e^{\theta} - 1 \right )^m = \sum_{n=m}^{\infty} 
{\cal S}_n^{(m)} {{{\theta}^n}\over{n!}} \quad . \eqno{(2.16)}
$$
${\cal D}_q$ satisfies, together with $z$ and $z {d\over {dz}}$, the 
commutation relations: 
$$
\bigl[ {\cal D}_q , z \bigr] = q^{z {d\over {dz}}} \quad ,\quad  
\left [ z {d\over dz} , {\cal D}_q \right ] = - {\cal D}_q \quad ,\quad 
\left [ z {d\over dz} , z \right ] = z \quad , \eqno{(2.17)}
$$ 
which can be recognized as a realization of relations (2.6) in the space 
${\cal F}$, with the identification
$$ 
N \to z {d\over dz} \quad ,\quad {\hat a}_q \to z \quad ,\quad 
a_q \to {\cal D}_q \quad , \eqno{(2.18)}
$$
with ~${\hat a}_q = {\hat a}_{q=1} = a^\dagger$~ and ~$\lim_{q\to1} a_q = a$ 
on ${\cal F}$. Let us stress that, while (2.17) are restricted to ${\cal F}$,
the operators (2.18) are related to the true algebraic structure (2.3-4). This
must be carefully kept in mind when considering their action over ${\cal F}
\otimes {\cal F}$. 
 
The relations analogous to (2.10) for the quantum case are
$$
{\hat a}_q u_n (z) ~=~ \sqrt{n + 1} ~u_{n+1} (z) \quad ,\quad 
a_q u_n (z) ~=~ q^{{n-1}\over {2}} 
{{[n]_q}\over {\sqrt{n}}} ~u_{n-1} (z) \; .
\eqno{(2.19)}
$$
 
There follows that the $q$-commutator  $[ a_q, {\hat a}_q ]$ is defined on 
the whole ${\cal F}$, where it acts as  
$$ 
[ a_q, {\hat a}_q ] f(z) = q^{N} f(z) = f(q z) \quad . 
\eqno{(2.20)}
$$
 
Because quantization of the algebra has been essentially obtained by replacing
the customary derivative with the finite difference operator, the above
discussion suggests$^{\, [15]}$ that whenever one deals with some finite scale
({\sl e.g.} with some discrete structure, lattice or periodic system) which
cannot be reduced to the continuum by a limiting procedure, then a deformation
of the operator algebra acting in ${\cal F}$ should arise. Deformation of the
operator algebra is also expected whenever the system under study involves
periodic (analytic) functions, since periodicity is but a special invariance
under finite difference operators. 
 
We finally note that eq. (2.20) provides a remarkable result since it shows
that the action of the $q$-WH algebra commutator $[ a_q, {\hat a}_q ]$, which
is a {\it linear} form in $a_q$ and ${\hat a}_q$, may be represented in the FBR
as the action of the operator $q^N$ which is {\it nonlinear} in the FBR
operators $a$ and $a^\dagger$. 
 
In the sections to follow, we shall apply such a result as well as the concepts
discussed above to several different cases of physical interest.

\bigskip
\noindent {\bf 3. $q$-Weyl-Heisenberg algebra and coherent states} 
\bigskip
 
In this section we shall discuss the connection between the $q$-WH algebra and
the CS formalism, exploiting the theory of entire analytic functions. More
specifically, as an application of the result expressed by eq.(2.20), we shall
show that the action of the commutator $[ a_{q} , {\hat a}_{q} ]$ may be
related in the FBR to the action of the CS displacement operator.

The Fock--Bargmann representation provides a simple and transparent frame to
describe the usual CS$^{\, [12],[16]}$, which are written in the form: 
$$ 
|\alpha > = {\cal D}(\alpha ) |0> \quad ;\quad     
a |\alpha> = \alpha |\alpha>\quad , \quad a |0> = 0\quad , \quad   
\alpha \in \CcC  \quad , \eqno{(3.1)}
$$
$$
|\alpha> = \exp\biggl({-|\alpha|^2\over 2} \biggr) \sum_{n=0}^\infty {{\alpha
^n}\over {\sqrt{n!}}} |n> = \exp\biggl({{-|\alpha|^2}\over 2}\biggr)
\sum_{n=0}^\infty u_n(\alpha) |n> \quad . \eqno{(3.2)}
$$ 
 
The relation between the CS and the entire analytic function basis ~$\{ u_n(z)
\}$ (eq. (2.8)) is here made explicit: $u_n (\alpha) = {\rm e}^{{1\over
2}|\alpha|^2} <n|\alpha>$.  The unitary displacement operator ${\cal
D}(\alpha)$ in (3.1) is given by: 
$$
{\cal D}(\alpha) = \exp\bigl(\alpha a^\dagger -{\bar \alpha} a \bigr)
= \exp\biggl(-{{|\alpha|^2}\over 2}\biggr) \exp\bigl(\alpha a^\dagger\bigr)
\exp\bigl(-{\bar \alpha} ~a\bigr)  \; , \eqno{(3.3)} 
$$
for which the following relations hold
$$
{\cal D}(\alpha) {\cal D}(\beta) = \exp\bigl( i {\it Im}(\alpha {\bar
\beta})\bigr) {\cal D}(\alpha + \beta) \quad , \eqno{(3.4)} 
$$ 
$$ {\cal D}(\alpha) {\cal D}(\beta) = \exp\bigl( 2 i {\it Im}(\alpha {\bar
\beta})\bigr) {\cal D}(\beta) {\cal D}(\alpha) \quad . \eqno{(3.5)} 
$$
 
Eq. (3.4) is nothing but the WH group law, also referred to as the Weyl
integral representation. 
 
It is well known that, in order to extract a complete set of CS   $\{
|\alpha_{\bf n}> \}$ , from the overcomplete set $\{ |\alpha >\}$ it is
necessary to introduce a regular lattice $L$ in the $\alpha$-complex plane$^{\,
[11]}$. The points (lattice vectors) $\alpha_{\bf n}$ of $L$ ~$\bigl(\{
\alpha_{\bf n}\in \CcC \, ; \,  {\bf n} = (n_1 , n_2)\, ; \, n_j \in \ZzZ \}
\bigr)$ are given by $\alpha_{\bf n} = n_1 \Omega_1 + n_2 \Omega_2 \equiv {\bf
n} \cdot {\bf \Omega}$, with the two lattice periods $\Omega_j \, , \, j=1,2$
linearly independent, {\sl i.e.} such that ${\rm Im}\, \Omega_1 {\bar \Omega_2}
\not= 0$. 
 
By closely following ref. [12], we recall that the set $\{ |\alpha_{\bf n}>\}$
(with exclusion of the vacuum state $|0\!  > \equiv |\alpha_{\bf 0}\! >$) can
be shown to be complete, invoking square integrability along with
analyticity$^{\,[17]}$, if the lattice elementary cell has area ${\rm Im}\,
\Omega_1 {\bar \Omega_2} = \pi$ ($L$ is called, in this case, the von Neumann
lattice). 
 
The lattice vectors $\alpha_{\bf n}$ describe the discrete translation 
invariance of $L$: 
$$
\alpha_{{\bf n} + {\bf m}} = \alpha_{\bf n} + \alpha_{\bf m} \quad ,  
\eqno{(3.6)}
$$
namely, 
$$
{\rm e}^{{\alpha_{\bf n}} 
{d\over {d\alpha}}} |\alpha >\big |_{\alpha = 
\alpha_{\bf m}}  = | \alpha_{{\bf n} + {\bf m}} > \quad . \eqno{(3.7)} 
$$ 
 
We now map the denumerable set of points $\alpha_{\bf n}$ onto the set $\{
z_{\bf n} ~ | z_{\bf n}\in \CcC \}$ with $\displaystyle{z_{\bf n} \equiv {\rm
e}^{\alpha_{\bf n}}}$. Assuming that the two periods $\Omega_1$ and $\Omega_2$
have imaginary parts incommensurate with $\pi$ and among themselves, such a map
is one-to-one and no point $z_{\bf n}$ lies on the real axis in the $z$ plane~
(notice that the set $\{ z_{\bf n} \}$ does $\underline{\rm not}$ constitute a
lattice in $z$, but it has the structure of concentric circles of radius 
${\rm e}^{{\rm Re}\alpha_{\bf n}}$). 
 
On the other hand, the function $z = {\rm e}^{\alpha}$, which most naturally
interpolates among these points, is analytical in its domain of definition, and
one may introduce, along with the basis functions $\{u_n(\alpha) \}$ the new
functions $\{ \tilde u_n(z) \equiv u_n(\ln z) = u_n(\alpha)\}$, with $\tilde
u_n(z) \in  {\cal F}$. 
 
Introducing the complex parameter $\displaystyle{q_{\bf m} \equiv {\rm
e}^{\alpha_{\bf m}} }$, it is then straightforward to check that, in ${\cal
F}$~, 
$$
[ a_{q_{\bf m}} , {\hat a}_{q_{\bf m}} ]
{\tilde u}_n(z) =
{q_{\bf m}}^{z{d\over dz}}{\tilde u_n(z)} =
\tilde u_n{(q_{\bf m}z)} =
u_n(\alpha + \alpha_{\bf m}) =
{q_{\bf m}}^{d\over {d\alpha}}u_n(\alpha) \quad ,
\eqno(3.8)
$$
where we used eq. (2.20). We have, therefore, 
$$\eqalign{
[ a_{q_{\bf n}} , {\hat a}_{q_{\bf n}} ] {\tilde f_{m}(z)}
\big |_{z = z_{\bf r}}
&= \tilde f_{m}(q_{\bf n} z_{\bf r }) =
f_{m}(\alpha_{\bf r} + \alpha_{\bf n}) \cr 
= {q_{\bf n}}^{d \over{d\alpha}}
f_{m}(\alpha) \big |_{\alpha = \alpha_{\bf r}} 
&= \exp \bigl(-i Im \alpha_{\bf n} {\bar{\alpha}}_{\bf r} \bigr)
<m|{\cal D}(\alpha_{\bf n})|\alpha_{\bf r}> \quad, \cr}
\eqno{(3.9)}
$$
where we utilized eq. (3.4), together with the notation $\displaystyle{f_{m}
(\alpha) \equiv \exp\bigl(-{{|\alpha|^{2}}\over{2}} \bigr) u_{m}(\alpha)}$.
 
The operator $\displaystyle{[ a_q , {\hat a}_q ] \, \big |_{q = q_{\bf n}}}$,
which realizes the mapping $\tilde f(z_{\bf m})~  \mapsto \tilde f(q_{\bf n}
z_{\bf m})$, can, in such way, be thought of as extendible to the map on the
$\alpha$-plane  $| \alpha_{\bf m} > \mapsto | \alpha_{{\bf n} + {\bf m}} >$. 
 
We also observe that in the FBR we have, for any $f \in {\cal F}$
$$
{\cal D}(\beta) f(\alpha) = \exp\biggl
(-{{|\beta|^2}\over 2}\biggr) \exp(\alpha \beta)
f(\alpha - {\bar \beta}) \quad ,\quad f \in {\cal F} \quad , \eqno{(3.10)} 
$$
so that, in view of eq. (3.8), we can write, for $q={\rm e}^{\zeta}$ and, once
more, $z = {\rm e}^{\alpha}$, 
$$
[ a_q, {\hat a}_q ] {\tilde f(z)} = {\rm e}^{{1\over 2} |\zeta |^2} 
{\bar q}^{\alpha} {\cal D}(-{\bar \zeta}) f(\alpha ) 
\quad . \eqno{(3.11)}
$$
 
Eqs. (3.9) and (3.11) show the relation between the $q$-WH algebra operator
$\displaystyle{[ a_q , {\hat a}_q ]}$ and the CS displacement operator,
establishing a relation between the quantum algebra (2.6) (or equivalently, at
this level, (2.3)), in the frame of the theory of entire analytical functions,
and the theory of CS. 
 
Once more we are led to conclude that the existence of a quantum deformed
algebra signals the presence of finite lengths in the theory and provides the
natural framework for the physics of discretized systems, the $q$-deformation
parameter being related with the lattice constants. 
 
The lattice structure is also of crucial relevance in the relation between
theta functions and the complete system of CS. 
 
In order to establish such a relation, we look for the common eigenvectors
$|\theta>$ of the CS operators ${\cal D}(\alpha_{\bf n})$ associated to the
regular lattice $L$ $^{\, [12]}$. A common set of eigenvectors exists if and
only if all the ${\cal D}(\alpha_{\bf n})$ commute, {\it i.e.} when the ${\cal
D}(\Omega _j)$ commute, as indeed it happens on the von Neumann lattice (cf.
eq.(3.5)). 
 
The eigenstates $|\theta>$ are characterized by two real numbers $\epsilon _1$
and $\epsilon _2$, $|\theta> \equiv |\theta _\epsilon>$, eigenvectors of ${\cal
D}(\Omega _i)$: 
$$
{\cal D}(\Omega _j) |\theta_\epsilon> = e^{i \pi \epsilon _j}
|\theta_\epsilon>\quad , \quad j=1,2 \quad , \quad 0 \leq \epsilon _j \leq 2 
\quad . \eqno{(3.12)}
$$
In other words, $|\theta _\epsilon>$, which belongs to a space which is the
extension$^{\, [12]}$ of the Hilbert space where the operators ${\cal
D}(\alpha)$ act, corresponds to a point on the two-dimensional torus.  The
action of ${\cal D}(\alpha)$ on $ |\theta _\epsilon>$ generates a set of
generalized coherent states.  Use of eqs. (3.12) and (3.4) gives 
$$
{\cal D}(\alpha_{\bf m}) |\theta_\epsilon> =
 e^{i \pi F_\epsilon({\bf m})} |\theta _\epsilon> 
\quad , \eqno{(3.13)}
$$
with $\alpha_{\bf m} = {\bf m}\cdot {\bf \Omega}$, an arbitrary lattice vector, 
and
$$ 
F_\epsilon({\bf m}) = m_1 m_2 + m_1 \epsilon _1 + m_2 \epsilon _2 \quad . 
\eqno{(3.14)}
$$
On the other hand, the system of CS is associated, in the FBR, with a
corresponding set of entire analytic functions, say $\theta_{\epsilon}
(\alpha)$. Eq. (3.10) with $\bar \alpha ~= -\alpha_{\bf m}$ shows that eq.
(3.13) may be written  as 
$$
\theta_\epsilon (\alpha+\alpha_{\bf m}) = \exp\bigl({i \pi F_\epsilon
(-{\bf m})}\bigr) \exp
\biggl({{|\alpha_{\bf m}|^2}\over 2}\biggr) \exp({\bar \alpha}_{\bf m} 
\alpha) \theta_\epsilon(\alpha) \quad ,
\eqno{(3.15)}
$$ 
which is the functional equation for the theta functions$^{\, [12],[18],[19]}$. 
We emphasize that such relation is obtained by considering the CS system
corresponding to the von Neumann lattice $L$. The relation with the $q$-WH
algebra is established by realizing that in $\cal F$ the functional equation
(3.15) reads 
$$
[ a_{q_{\bf m}} , {\bar a}_{q_{\bf m}} ] {\tilde \theta(z)} =
{\tilde \theta(q_{\bf m}z)} =
\exp\bigl({i \pi F_\epsilon
(-{\bf m})}\bigr) \exp
\biggl({{|\alpha_{\bf m}|^2}\over 2}\biggr) \exp({\bar \alpha}_{\bf m} 
\alpha) \theta_\epsilon(\alpha) \; .
\eqno{(3.16)}
$$ 
 
It is interesting to observe that the commutator $[ a_q, {\hat a}_q ]$  acts as
shift operator on the Von Neumann lattice whereas it acts as $z$-dilatation
operator ~$(z \to q z)$~ in the space of entire analytic functions or, else, as
the $U(1)$ generator of phase variations in the $z$-plane, ${\rm arg}(z)~\to 
{\rm arg}(z) + \theta $, when $q = {\rm e}^{i\theta}$. 
 
It is remarkable that eqs. (3.8), (3.9) and (3.11) represent the action of the
$q$-WH algebra commutator $[ a_q, {\hat a}_q ]$, (bi-)linear in $a_q$ and
${\hat a}_q$, through the action of the CS displacement operator which is
nonlinear in the FBR operators $a$ and $a^\dagger$. Conversely, the nonlinear
operator ${\cal D}(\alpha)$ is represented by the linear form ~$[ a_q , {\hat
a}_q ]$ in the $q$-WH algebra. 
 
\bigskip
\noindent {\bf 4. $q$-Weyl-Heisenberg algebra and the squeezing generator}
\bigskip
 
Over a Hilbert space of states identified with the space  ~${\cal F}$ of entire
analytic functions $\psi (z)$, the identity 
$$
2 z {d\over {dz}} \psi (z) = \left\{{1\over 2} 
\left[\left(z + {d\over {dz}}\right)^2 -
\left(z - {d\over {dz}}\right)^2\right] - 1\right\} \psi (z)\; , \eqno{(4.1)}
$$
holds. In the present section we set $z \equiv x + iy$, $x$ denoting the 
position coordinate variable.
For convenience, we introduce the operators 
$$ 
{\alpha} = {1\over {\sqrt{2}}} \bigl( z +  {d\over {dz}}
\bigr) \quad ,\quad { \alpha}^\dagger = {1\over {\sqrt{2}}} \bigl( z -
{d\over {dz}}\bigr)\quad , \quad 
[\alpha, \alpha^\dagger ] = \IiI \quad ,
\eqno{(4.2a)}
$$
namely, in terms of the FBR operators $a$ and $a^\dagger$,
$$
z = {1\over {\sqrt{2}}} ( \alpha + {\alpha}^\dagger )~ \to a \quad , \quad 
{d\over {dz}} = {1\over {\sqrt{2}}} ( \alpha - {\alpha}^\dagger )
~\to a^\dagger \quad .
\eqno{(4.2b)}
$$
 
One should notice that, in ~${\cal F}$, $~{\alpha}^\dagger$ is indeed the
conjugate of $\alpha$, as thoroughly discussed in Ref. [10] (see also [12]). We
observe that, in the limit $y \to 0$,~ $\alpha$ and $\alpha^\dagger$ turn into
the conventional annihilation and creator operators $a$ and $a^\dagger$
associated with $x$ and $p_x$ in the canonical configuration representation,
respectively. 
 
We shall now prove that the operator 
$$
[ a_q, {\hat a}_q] = {1\over{\sqrt q}}~\exp\biggl({\zeta \over 2}\bigl(\alpha^2
- {\alpha^\dagger}^2\bigr)\biggr) \equiv 
{1\over{\sqrt q}} {\hat {\cal S}(\zeta)} , \eqno{(4.3)} 
$$
where $q = e^\zeta$ (for simplicity, assumed to be real), which reminds the
squeezing operator of quantum optics $^{[20]}$, acts indeed in ~${\cal F}$ as
well as a squeezing operator in the configuration representation in the limit
$y \to 0$. 
 
We note first that, as it is well known$^{\, [20]}$, the right hand side of
(4.3) is an $SU(1,1)$ group element. In fact, by defining $K_{-} = {1\over
2}\alpha ^2$, $K_{+} = {1\over 2}\alpha^{\dagger 2}$, $K_{z} = {1\over
2}(\alpha^\dagger \alpha + {1\over 2})$, one easily checks they close the
algebra $su(1,1)$. 
 
We have also (see eq. (2.20)):
$$
\eqalign{
[ a_q, {\hat a}_q] \psi (z) = \exp\Bigl(\zeta &z {d\over dz}\Bigr) \psi (z) =
{1\over{\sqrt q}} \exp\Bigl({\zeta\over 2}\bigl(\alpha^2 -
{\alpha^\dagger}^2\bigr) \Bigr)\psi (z)\cr 
&= {1\over{\sqrt q}} {\hat {\cal S}}(\zeta) \psi (z) \equiv 
{1\over{\sqrt q}} \psi_{s}(z)\quad , \cr}
\eqno{(4.4a)} 
$$
with $\psi _s(z)$ the squeezed states in FBR, and in the $y \to 0$ limit (still
in ${\cal F}$) 
$$
\eqalign{
{\hat {\cal S}}^{-1}(\zeta) \alpha {\hat {\cal S}}(\zeta) &\to  
{\hat {\cal S}}^{-1}(\zeta) a {\hat {\cal S}}(\zeta) \cr
{\hat {\cal S}}^{-1}(\zeta) {\alpha}^\dagger {\hat {\cal S}}(\zeta) &\to
{\hat {\cal S}}^{-1}(\zeta) a^\dagger {\hat {\cal S}}(\zeta) \cr
{\hat {\cal S}}^{-1}(\zeta) z {\hat {\cal S}}(\zeta) = {1\over{q}}z \to 
{1\over{q}}x \quad &, \quad 
{\hat {\cal S}}^{-1}(\zeta)p_{z} {\hat {\cal S}}(\zeta) = q p_{z} \to 
qp_x \quad ,\cr}
\eqno{(4.4b)}
$$
where $\displaystyle{p_z \equiv -i{d\over{dz}}}$. We thus see that in the limit
$y \to 0$ 
$$
\eqalign{
&\int d\mu(z) {\bar \psi}(z){\hat {\cal S}}^{- 1}(\zeta) z {\hat {\cal S}}
(\zeta) \psi (z)   \to {1\over{q}}<x> \cr
&
\int d\mu(z) {\bar \psi}(z){\hat {\cal S}}^{- 1}(\zeta) p_{z} {\hat {\cal S}}
(\zeta) \psi (z)   \to  q<p_{x}> \cr}
\eqno{(4.5)}
$$
so that the root mean square deviations $\Delta x$ and $\Delta p$ of position
and momentum satisfy 
$$
\Delta x \Delta p = {1\over 2} \quad ,\quad 
\Delta{x} = {1\over q} {\sqrt{1\over 2}}\quad ,\quad \Delta {p} = 
q{\sqrt{1\over 2}}\quad . 
\eqno{(4.6)}
$$
Manifestly the $q$-deformation parameter plays here the r\^ole of squeezing
parameter and the commutator $[ a_q, {\hat a}_q ]$ acts, up to a numerical
factor, like the conventional squeezing generator with respect to the operators
$\alpha$ and $\alpha^\dagger$. 
 
Let us finally observe that, in view of the holomorphy conditions holding for 
$f(z)\in {\cal F}$ 
$$
{d\over dz} f(z) = {d\over dx} f(z) = -i {d\over dy} f(z) \quad ,\eqno{(4.7)}
$$
one finds 
$$
z {d\over dz} = z {d\over dx} = x {d\over dx} + iy{d\over dx} =
-ix {d\over dy} + iy{d\over dx} = x p_y - y p_x \equiv L \quad , \eqno{(4.8)}
$$
with $L$ an angular momentum operator.  This implies that, for $\zeta = i
\theta$, with $\theta$ real, the commutator $[ a_q, {\hat a}_q ]$ acts in
${\cal F}$ as the $U(1)$ group element 
$$
[ a_q, {\hat a}_q ] = e^{i \theta L} \quad . 
\eqno{(4.9)}
$$
 
\bigskip 
\noindent {\bf 5. Lattice  Quantum Mechanics and Bloch functions.}
\bigskip
 
In this section we first recall  the structure of Lattice Quantum Mechanics
(LQM) and then relate it to $q$-WH. For completeness we also construct lattice
CS minimizing the lattice position-momentum uncertainty relation. Finally, we
close with an application to Bloch functions. 
 
We limit ourselves to 1-dimensional lattice: the extension to more dimensions
requires the accurate use of the coproduct in the quantum-algebraic scheme, and
we do not extend our analysis to such a case in the present paper. 
 
A 1-dimensional lattice quantum system is defined on the configurational
Hilbert space ${\cal G} = l^2(\epsilon \ZzZ )$, where $\ZzZ$ denotes the set of
integers $n$ and $\epsilon$ is the lattice spacing. The (hermitian) position
operator, ${\hat x}_{\epsilon}$, is defined by 
$$
[{\hat x}_{\epsilon} f](x_{n}) = x_{n} f(x_{n}) = \epsilon n f(x_{n}) 
\quad , \quad f \in {\cal G} \quad,  \eqno{(5.1)}
$$
whereas the (hermitian) lattice momentum operator ${\hat p}_{\epsilon}$ is
defined by 
$$
[{\hat p}_{\epsilon} f](x_{n}) = -i [{\cal D}_{\epsilon} f](x_{n})\quad ,
\eqno (5.2)
$$
where ${\cal D}_{\epsilon}$ is the symmetrized, finite difference gradient 
$$
[{\cal D}_{\epsilon} f](x_{n}) = (2 \epsilon)^{-1} 
[f(x_{n+1}) - f(x_{n-1})] 
\quad . \eqno (5.3)
$$
In this section we use the finite difference operator ${\cal D}_{\epsilon}$,
which is slightly different from ${\cal D}_{q}$, to follow the conventional use
in the literature on LQM. The relation between the two operators is manifest: 
${\cal D}_{\epsilon} = {1\over 2} \bigl ( {\cal D}_{q_n} + {\cal D}_{q_{n-1}} 
\bigr )$, with $q_n = 1+ n^{-1}$. 
 
The dual momentum-space representation of the above operators is  simply 
$$
[{\hat x}_{\epsilon} f](k) = i {d\over {dk}} f(k)\quad , 
\eqno (5.4a)
$$
$$
[{\hat p}_{\epsilon} f](k) = \epsilon^{-1} {\sin (k \epsilon)}
f(k) \quad , \eqno (5.4b)
$$
respectively, where $f(k)$ is the Fourier conjugate of $f(x_n)$, $k$ belonging
to the first Brillouin zone (BZ), $|k| \leq \pi/\epsilon$. 
 
Over ${\cal G}$ one has the following commutation relation between position and
momentum 
$$
\bigl[ [{\hat x}_{\epsilon} , {\hat p}_{\epsilon} ] f \bigr](x_{n}) =
{i\over 2} \left (f(x_{n+1}) + f(x_{n-1})\right ) \equiv
i [{\hat P}_\epsilon f] (x_n) \quad . \eqno (5.5)
$$
The operator ${\hat P}_\epsilon$ has the following explicit expression.
Consider the set of entire functions $f(x)$ whose restriction to the lattice
points $x_n$ is the set $\{f(x_n)\}$: then  ${\hat P}_\epsilon$ is given by 
${\hat P}_\epsilon = \cos (\epsilon  {\hat p} )\;$ , with ${\hat p} = -i
{d\over {dx}}$, as it follows from 
$$
f(x_{n \pm 1}) \equiv f(x_{n} \pm \epsilon) =
\bigl[\exp(\pm i \epsilon {\hat p}) f \bigr](x_{n}) \quad . 
\eqno (5.6)
$$
 
The operators ${\hat P}_\epsilon$, $\hat p_\epsilon$  and ${\hat x}_{\epsilon}$
generate the algebra $E(2)$: 
$$
[{\hat x}_{\epsilon} , {\hat p}_{\epsilon} ] = i {\hat P}_\epsilon ~~,~~ [{\hat
x}_{\epsilon} , {\hat P}_\epsilon ] = -i \epsilon^2 {\hat p}_{\epsilon} ~~,
~~ [{\hat P}_\epsilon , {\hat p}_{\epsilon} ] = 0  \quad ,  \eqno (5.7)
$$
which contracts to the Weyl--Heisenberg algebra in the limit $\epsilon
\rightarrow 0$: the discrete lattice spacing $\epsilon$ plays thus the r\^ole
of deformation parameter. 
 
Upon introducing, in momentum space, the angle $\phi = k \epsilon$, $-\pi \leq
\phi \leq \pi$, ranging on the whole unit circle, we set 
$$
L_3 = -i {d\over {d\phi}} = - {i\over \epsilon} {d\over {dk}} \quad ,\quad 
L_1 = \cos \phi \quad ,\quad L_2 = \sin \phi \quad .  \eqno{(5.8)} 
$$
The algebra (5.6) takes then the more customary $E(2)$ form
$$
[ L_3, L_1 ] = i L_2 \quad ,\quad  [ L_3, L_2 ] = -i L_1 \quad ,\quad  
[ L_1, L_2 ] = 0 \quad . \eqno{(5.9)}
$$
It is interesting to observe that eqs. (5.8) and (5.9) correspond to the ones
reported in ref. [21] in the framework of a discussion on the phase-angle
commutation relations. 
 
In order to show that the structure underlying LQM is, once more, $q$-WH,
we introduce now -- as in sect. 3 -- a
conformal image $\tilde {\cal F}$ of the configurational Hilbert space ${\cal
G}$ in the following way. 
We map the denumerable set of lattice points 
$x_{n} = {\epsilon}n$ onto the set 
of points $z_n \equiv {\rm e}^{ix_n}$ on the unit circle.
Assuming that $\eta = {{2\pi}\over{\epsilon}}$ is an irrational number,
such a set $\{z_n\}$ is dense on the circle, $z_n\neq z_m$ for $n \neq m$ and the 
number of lattice points mapped on each single image of the 
circle is the maximum 
integer less than $\eta$. 
 The function $z = {\rm e}^{ix}$ (not to
be confused with the $z$ variable introduced in sect. 4), which most naturally
interpolates among these points, is analytical in its domain of definition.
One may therefore introduce   
the function $\tilde f(z)
\equiv f(-i \log z) = f(x)$, provided the bridging of the cut 
is realized by the customary analytic continuation of the
{\sl log} function. We have 

$$
i [{\hat p} f](x) = {d \over {dx}} f(x) = 
i z {d \over {dz}}
f(-i\log z)\; = i z {d \over {dz}} {\tilde
f}(z)\quad ,\quad z = e^{ix}\quad ,\quad {\tilde f} \in
{\tilde {\cal F}}\; . \eqno (5.10) 
$$ 

Once more,
we restrict the space ${\cal F}$ to the space of
the functions $\{f(x_n)\,| f \in {\cal F}\}$ over the lattice points $x_n$. 
 
Upon noticing that the operators $A^\dagger = e^{ix}$, $A = -i e^{-ix}
{d\over {dx}}$, $N = -i {d\over {dx}} = {\hat p}$ provide a realization of
WH-algebra, the form of $A^\dagger$ suggests that the same realization of FBR
may be adopted in ${\tilde {\cal F}}\left(\{f(x_n)\}\right)$ for the LQM
as well. Eq. (5.6) now implies
$$
f(x_{n+1}) = [e^{i \epsilon {\hat p}} f](x_{n}) = {q^N} {\tilde f}(z_{n}) =
{\tilde f}(q z_{n}) = {\tilde f}(z_{n+1})\quad , \eqno (5.11) 
$$
where we have set $e^{i \epsilon}=q$, and used eq. (2.20).
 
This makes clear that the algebraic structure of LQM is intimately related with
the $q$-WH algebra, the $q$-deformation parameter being determined by the
discrete lattice length $\epsilon = -i \log \, q$. 
 
The same conclusion can be reached by constructing LQM in momentum space. In
such a case one has to consider a conformal image $\tilde {\cal H}$ of the
Hilbert space in the momentum representation, by setting  $ z = e^{i \phi}\;$, 
so that (cf. eq.(5.8)) 
$$
-i {d \over {d \phi}} = -{i\over \epsilon} {d\over {dk}} = z {d \over {dz}}
\quad .\eqno (5.12) 
$$
 
On the other hand, 
$$
L_{3} f(\phi) = -i {d \over {d \phi}} f(\phi) = z {d \over {dz}} {\tilde f}(z)
= N {\tilde f}(z) \quad ,\quad {\tilde f} \in {\tilde {\cal H}} \quad , \eqno
(5.13) 
$$
and eq. (2.20) gives:
$$
[a_{q},{\hat a}_{q}]{\tilde f}(z) =
q^N {\tilde f}(z) = {\tilde f} (qz) = 
f(\phi+\epsilon) = e^{i \epsilon L_{3}} f(\phi) \quad . \eqno (5.14)
$$
The $E(2)$ algebra (5.9) is now realized in the momentum space FBR as 
$$
\eqalign{
\left [ [L_{1} , L_{3}] {\tilde f}\right ] (z) = -i \left [ L_{2} {\tilde f} 
\right ] (z)\quad &, \quad  
\left [ [L_{2} , L_{3}] {\tilde f}\right ] (z) = i \left [ L_{1} {\tilde f} 
\right ] (z)\quad , \cr    
\left [ [L_{1} , L_{2}] {\tilde f}\right ] (z) &= 0 \quad ,\cr} \eqno (5.15)
$$
with ${\tilde f} \in {\tilde  {\cal H}}$, and the identifications (see
eq.(5.8)) 
$$
L_{1} ={ {z+{\bar z}} \over 2}\; ,\; L_{2} = {{z-{\bar z}} \over {2i}}\; ,
\; L_{3} = z {d \over {dz}}\; ,\; L_{+} = z\; , \; L_{-} = {\bar z}\; .
\eqno (5.16)$$
 
In conclusion, in this representation $[a_{q}, {\hat a}_{q}]$ is nothing but
the $e^{i \epsilon L_{3}}$ group  element of $E(2)$. We  also  note that $z^{n}
= e^{i n \phi}$, $n$ integer, is the eigenfunction of $L_{3}$ associated with
the integer eigenvalue $n$ of the  number operator in the FBR 
$$
L_{3} z^{n} = N z^{n} = n z^{n}\quad . \eqno (5.17)
$$
 
We can now construct the CS minimizing the lattice position-momentum
uncertainty relation. To begin with, by following standard procedures$^{\,
[21],[23]}$, we get, in the LQM scheme above, the uncertainty inequalities 
$$
\eqalign{
\Delta^{\!2} \left({\hat x}_{\epsilon} \right)
\Delta^{\!2}  \left({\hat p}_{\epsilon} \right) &\geq
{1 \over 4} {\langle \cos(k \epsilon) \rangle}^{2} \quad ,\cr 
\Delta^{\!2} \left({\hat x}_{\epsilon}\right)
 \Delta^{\!2} \left(\cos(k \epsilon) \right) &\geq 
{1 \over 4}{\epsilon^{2}} {\langle {\sin(k \epsilon)}
\rangle}^{2} \quad , \cr} \eqno (5.18)
$$
where $\langle {\hat A} \rangle = \int dk \Psi^{*}(k) {\hat A} \Psi (k)\;  $
denotes quantum expectation on the lattice and $ \Delta^{\!2}({\hat A}) =
\langle {\hat A}^2 \rangle - {\langle {\hat A} \rangle}^2 $ the (square)
variance. Relations (5.18) are written in momentum space for convenience, and
are consequences of eqs. (5.7). The continuum limit $\epsilon \rightarrow 0$
corresponds to opening the circle into a line. In $d$ dimensions the limit 
$\epsilon \rightarrow 0$, would be an isometric and conformal mapping of the
torus on the plane (decompactification)$^{\, [22]}$. The states minimizing the
uncertainty products of eqs.(5.18) must satisfy$^{[21],[23]}$ 
$$
\left({\hat x}_{\epsilon} + i \gamma {\hat p}_{\epsilon} \right) \Psi = \lambda
\Psi \quad , \eqno (5.19)
$$
where $\lambda = \langle {\hat x}_{\epsilon} \rangle + i \gamma \langle {\hat
p}_{\epsilon} \rangle$ , and $\gamma$ is connected with the position and
momentum variances $\Delta\left({\hat x}_{\epsilon} \right)$~,~$ \Delta
\left({\hat p}_{\epsilon} \right)$. Relation (5.19) becomes, in momentum space,
$$
\left[{d \over{d(\epsilon k)}} + {\bar \gamma} \sin (\epsilon k)\right] 
\Psi(k) = -i {\bar \lambda} \Psi(k)\quad , \eqno (5.20)
$$
where ${\bar \lambda} = \lambda \epsilon^{-1} \,, \, {\bar \gamma} = \gamma
\epsilon^{-2}$. Its solution is 
$$
\Psi(k) = G \exp \left [{\bar \gamma} \cos (\epsilon k) - i {\bar \lambda}
\epsilon k \right ]\quad , \eqno (5.21)
$$
where the normalization constant $G$ is given by $G = 2\pi {\epsilon}^{-1} 
I_{0}(2{\bar \gamma})$ , $I_{0}$ denoting the modified Bessel function of the
first kind of order $0$. Notice that, in the continuum limit $\epsilon \to 0$,
the Fourier transform ${\tilde \Psi}(x)$ of (5.21) becomes 
$$
\tilde\Psi (x) = \left({\gamma \pi}\right)^{-{1\over 4}}\, \exp \left\{- \left
[ (2\gamma)^{-1} (x - \langle {\hat x} \rangle)^2 + i \langle {\hat p} \rangle
(x - \langle {\hat x} \rangle)\right ] \right\} \quad , \eqno{(5.22)}
$$
which is just the minimum uncertainty wavefunction given by Schr\"odinger$^{\,
[24]}$. By setting $z = <x> + i \gamma <p>$, eq. (5.22) defines the usual
coherent states$^{\, [12]}$. The wave-function (5.21) is the coherent state for
a system with discretized position and momentum or, equivalently, endowed with
some periodic constraint. 
 
Finally, we close this section by showing, as a further application of eq.
(2.20), that Bloch functions provide a representation of $q$-WH algebra. The
basic observation is that the  functions $z = e^{i \phi}$ and $z^{n}$ play also
a crucial r\^ole in the Bloch function theory$^{[25]}$. Bloch theorem indeed
states the existence, in the presence of a periodic potential $V(x_n ) = V(x_n
+ \epsilon )$, of solutions of the related Schr\"odinger equation of the form 
$$
\psi(x_{n}) = e^{\pm i k x_{n}} v_{k}(x_{n})\quad , \eqno (5.23)
$$
with $v_{k}(x_{n}) = v_{k}(x_{n} + \epsilon)$. $\psi(x_{n})$ is referred to as
a Bloch function. Limiting ourselves for simplicity to considering the plus
sign in (5.23), $\psi(x_{n})$ has the property 
$$
\psi(x_{n} + \epsilon) = e^{i k \epsilon} \psi(x_{n}) =
z \psi(x_{n})\quad . \eqno (5.24)
$$
We  thus see that the choice of the variable $z= e^{i k \epsilon}$ turns out to
be natural in the case of periodic potentials: 
$$
\psi(x_{n}) = z^{n} v_{k}(x_{n})\quad ,\quad 
\psi(x_{n} + \epsilon) = z^{n+1} v_{k}(x_{n})\quad . \eqno (5.25)
$$
Since $z^{n} = (z_{n})^k$, from eq. (5.11)
$$
q^N (z_{n})^k = (q z_{n})^k = (e^{i \epsilon} e^{i x_{n}})^k =
e^{i k \epsilon (n+1)} = z^{n+1}\quad , \eqno (5.26)
$$
where $q^N$ is understood as defined on $\tilde {\cal F}$. We  thus have in
$\tilde {\cal F}$, by using eq. (2.20), 
$$
\psi(x_{n} + \epsilon) = 
\bigl[ [_{q} , {\hat a}_{q}] (z_{n})^k u_{k} \bigr](x_{n}) =
\bigl[ [_{q} , {\hat a}_{q}] \psi \bigr](x_{n})\quad . \eqno (5.27)
$$
In other words, the condition implementing the Bloch function periodicity
features is realized by the same operator in the $q$-WH algebra acting, 
in the FBR, as dilatation.
 
\bigskip
\noindent {\bf 6. Conclusions}
\bigskip
 
In this paper we have considered the $q$-deformation of the WH algebra  in
connection with typical problems in QM: discrete and/or periodic systems,
coherent states, squeezing, lattice quantum mechanics, resorting essentially to
the tool of finite difference operators. The above examples have been proposed 
as physically relevant representative of systems where $q$-deformation plays a
r\^ole. Such a collection of applications is not only interesting on its own,
but also as a {\it laboratory} where problems characterized by the common
feature of a discrete structure are treated in the framework of (entire)
analytic functions theory. 
 
The underlying aim is the $''$phylosofical$''$ claim that, whenever one deals
with a quantum system defined on a countable set of degrees of freedom, then
one has to work in the space of the related analytical functions and the
structure of the operator algebra is that of a quantum algebra. In this sense,
the term {\it deformation} does not sound as the most appropriate, since
$q$-algebras provide an algorithm of wide physical application, and not an
exception for situations {\it deformed} with respect to a standard one.
Moreover, quantum algebras stucture amounts to much more than deformation, and
in particular we expect that -- in the case of $q$-WH -- both the graded
structure and the operation of co-product have to be taken into account. Such a
general philosophy appears to find very concrete preliminary support in the
results discussed in this paper. 
 
From the point of view of group theory , we used the well known  mapping of a
$q$-algebra into the universal  enveloping algebra of the corresponding Lie
structure; to be specific, the mapping of finite difference operators into
functions of differential operators, which can be indeed achieved only by
operating on $C^\infty$ functions. This was the main reason to work with FBR
and to introduce the conformal image of both configuration and momentum space
in the study of LQM. 
 
An interesting and natural development, namely the extension from QM to Quantum
Field Theory by considering the infinite volume limit of the lattice system,
leads to the parametrization of the unitarily inequivalent representations of
the canonical commutation relations by means of the deformation parameter
$q\,^{\,[27]}$. Different values of the lattice spacing are thus described by
inequivalent representations. In this framework  finite temperature and
dissipative systems$^{\,[26]}$ may find an appropriate unified description. 
 
The notion of coproduct, although an essential tool in the discussion of
$q$-algebras, has not explicitly entered so far our analysis, in that in this
paper we have mainly focused our attention on the links between the $q$-WH
algebra and several structures of physical interest in $1-d$ only. We are aware
that a deeper understanding of such links may be achieved by investigating the
r\^ole played by the whole structure of Hopf algebra of the $q$-WH, that we
expect should be essential in dealing with systems in higher dimension. This
will hold in particular when dealing with the generalized definition of theta
functions, where the notion of mapping class group (connected with the braiding
features of the transformations of the coherent states manifold) comes into
play. 
 
Work is under progress along these latter directions. 
 
\bigskip
\noindent {\bf Acknowledgements}
\bigskip
We gratefully acknowledge useful discussions with O.V.Man'ko, J.Katriel
and A.I.Solomon.
 
\vfill\eject 
 
\baselineskip= 16 pt
\nopagenumbers
 
\phantom{xxxxx}
\bigskip
 
\line{{\bf References.}\hfill}
\bigskip

\ii {[1]} Drinfeld V.G., Proc. ICM Berkeley, CA; A.M. Gleason, ed,; AMS, 
	   Providence, R.I., 1986, page 798.
\ii {[2]} Jimbo M., Int. J. of Mod. Phys. {\bf A4} (1989) 3759.
 
\ii {[3]} Manin Yu.I., {\it Quantum groups and Non-Commutative Geometry},
	   Centre de Recherches Math\'ematiques, Montreal, 1988.
 
\ii {[4]} Abe E., {\it Hopf algebras}, Cambridge tracts in Math. no. 74,
	   Cambridge Univ. Press, Cambridge, 1980.
 
\ii {[5]} E. Celeghini, R. Giachetti, E. Sorace, and M. Tarlini, J. Math. 
Phys. {\bf 31} (1990) 2548; and {\bf 32} (1991) 1155.

\ii {[6]}   L.C Biedenharn, J. Phys. {\bf 22} (1989) L873
 
\ii {[7]}   A.J. Macfarlane, J. Phys. {\bf 22} (1989) 4581
 
\ii {[8]}  P.P. Kulish, and N.Yu. Reshetikin, Lett. Math. Phys. {\bf 18} 
(1989) 143.

\ii {[9]}   E.Celeghini, T.D.Palev and M.Tarlini, Mod. Phys. Lett {\bf B5}
	(1991)187
 
\ii {[10]}   E. Celeghini, M. Rasetti and G. Vitiello, 
	     Phys. Rev. Lett. {\bf 66} (1991) 2056

\ii {[11]}   V.A. Fock, Z. Phys. {\bf 49} (1928) 339 
	
	V. Bargmann, Comm. Pure and Appl. Math. {\bf 14} (1961) 187
 
\ii {[12]}   A. Perelemov, {\it Generalized Coherent States and 
	Their Applications},  Springer-Verlag  
	Berlin 1986

\ii {[13]}   L.C. Biedenharn, and M.A. Lohe , Comm. Math. Phys. {\bf 146},
	  483 (1992)
 
\ii {\phantom{[13]}} F.H. Jackson, Messenger Math. {\bf 38}, 57 (1909)
 
\ii {\phantom{[13]}} T.H. Koornwinder, 
	Nederl. Acad. Wetensch. Proc. Ser. {\bf A92},
	97 (1989)        
 
\ii {\phantom{[13]}} D. I. Fivel, J. Phys. {\bf 24}, 3575 (1991)

\ii {[14]}   M. Abramowitz and I.A. Stegun, {\it Handbook of Mathematical 
	      Functions}, Dover Publications, Inc., New York, 1972 
 
\ii {[15]}   E. Celeghini, S. De Martino, S. De Siena, M. Rasetti and 
	      G. Vitiello, Mod. Phys. Lett. {\bf B7}, 1321 (1993)
 
\ii {[16]}   J.R. Klauder and B. Skagerstam, {\it Coherent States},
	      World Scientific, Singapore 1985

\ii {[17]}  V. Bargmann, P. Butera, L. Girardello and J.R. Klauder, Rep. on Math.
	Phys. {\bf 2} (1971) 221
 
\ii {[18]}  D. Mumford, {\it Tata Lectures on Theta, III}, Birkh\"auser Boston,
	Inc., 1991
 
\ii {[19]}  R. Bellman, {\it A brief introduction to theta functions},
	  Holt, Rinehart and Winston, N.Y. 1961

\ii {[20]}  H.P. Yuen, Phys. Rev. {\bf 13} (1976) 2226
	
\ii {\phantom{[20]}} E. Celeghini, M. Rasetti, M. Tarlini and G. Vitiello, Mod. 
	Phys. Lett. {\bf B3} (1989) 1213
 
\ii {[21]}  P. Carruthers and M. Nieto, Rev. Mod. Phys. {\bf 40} (1968) 411
 
\ii {[22]}  C. De Concini and G. Vitiello, Nucl. Phys. {\bf B116} (1976) 141
 
\ii {[23]}  R. Jackiw, J. Math. Phys. {\bf 9}, (1968) 339  
 
\ii {[24]}  E. Schr\"odinger, Naturwissenschaften {\bf 14} (1926) 664

\ii {[25]}  A.J. Dekker, {\it Solid State Physics}, Prentice-Hall, Englewood 
	  Cliffs 1957        
   
\ii {[26]}  E. Celeghini, M. Rasetti and G. Vitiello, Annals of Phys.(N.Y.)
	  {\bf 215} (1992) 156 
 
\ii {[27]} A. Iorio and G. Vitiello, Quantum groups and Von Neumann 
	      theorem, Mod. Phys. Lett. B, in press.
 
\vfill\eject
\bye